\documentclass[onecolumn,12pt,showpacs,preprintnumbers,amsmath,amssymb]{revtex4}

\usepackage{verbatim}% comments, among others
\usepackage{graphicx}% Include figure files
\usepackage{dcolumn}% Align table columns on decimal point
\usepackage{bm}% bold math

\begin{document}

\newcommand{\upd}{\textrm{d}}

\title{Interface Equations for Capillary Rise in Random Environment}

\author{T. Laurila}

\author{C. Tong}
\altaffiliation[currently at: ]{Department of Materials Science and
Engineering, McMaster University, Hamilton, Ontario, Canada}

\author{S. Majaniemi}
\altaffiliation[currently at: ]{Department of Physics, McGill University, 3600
University Street, Montr\'eal, QC, Canada H3A 2T8}

\author{T. Ala-Nissila}
\altaffiliation[also at: ]{Department of Physics, Brown University,
Providence RI 02912-8143}

\affiliation{Laboratory of Physics, P.O. Box 1100,
Helsinki University of Technology, FIN--02015 HUT }

\begin{abstract}
We consider the influence of quenched noise upon interface dynamics
in 2D and 3D capillary rise with rough walls by using phase-field
approach, where the local conservation of mass in the bulk is
explicitly included. In the 2D case the disorder is assumed to be in
the effective mobility coefficient, while in the 3D case we
explicitly consider the influence of locally fluctuating geometry
along a solid wall using a generalized curvilinear coordinate
transformation. To obtain the equations of motion for meniscus and
contact lines, we develop a systematic projection formalism which
allows inclusion of disorder. Using this formalism, we derive
linearized equations of motion for the meniscus and contact line
variables, which become local in the Fourier space representation.
These dispersion relations contain effective noise that is linearly
proportional to the velocity. The deterministic parts of our
dispersion relations agree with results obtained from other similar
studies in the proper limits. However, the forms of the noise terms
derived here are quantitatively different from the other studies.
\end{abstract}

%Short title:  Fluctuating Wall


\maketitle

\section{Introduction}

The dynamics and roughening of moving interfaces in disordered media
has been a subject of great interest in non-equilibrium statistical
physics since the 1980s. Relevant examples of physically and
technologically important processes include thin film
deposition~\cite{moldovan00}, fluid invasion in porous
media~\cite{bouchaud90,scheidegger57,krug97}, and wetting and
propagation of contact lines between phase
boundaries~\cite{moulinet04,joanny90,ertas94}. The understanding of
the underlying physics involved in interface roughening is crucial
to the control and optimization of these processes. Significant
progress in the theoretical study of interface dynamics has been
made and a number of theories have been developed~\cite{barabasi95},
which in some selected cases agree well with the experimental
findings \cite{paperburning}. Most of the theoretical understanding
in this field is based on modeling interface roughening with a local
stochastic equation of motion for the single-valued height variable
of the interface. However, there are several cases of interest where
such an approach cannot be justified {\it e.g.} due to conservation
laws in the bulk.
%
%The correlation functions of interface height typically
%show power law behavior in space and time, characterized by
%scaling exponents $\chi$ (roughness exponent) , $\beta$
%(growth exponent) and $z$ (dynamic exponent).
%
%the global and local roughness exponents differ~\cite{AMDHA}. Based on
%the scaling of structure factor, Ramasco {\it{et al.}}~\cite{RLR}
%proposed a generalized scaling theory and predicted the
%existence of a new class of growth models with novel
%anomalous scaling properties.
%
This is especially true for processes such as fluid invasion in
porous media, which is often experimentally studied by Hele-Shaw
cells~\cite{soriano02_03,paune03,geromichalos02} or imbibition of
paper~\cite{buldryev92,horvath95,amaral94}. It has been shown that
in such cases spatially local theories cannot provide a complete description
of the underlying dynamics. For describing the diffusive invasion
dynamics in such systems, a phase-field model explicitly including
the local liquid bulk mass conservation law has been developed
and applied to the dynamics of 1D imbibition fronts in
paper~\cite{Dube99_01}. This was achieved by a generalized
Cahn-Hilliard equation with suitable boundary conditions, which
couple the system to the reservoir. Numerical results for roughening
from the model are in good agreement with relevant experiments
\cite{laurila05}.
%Numerical results for spontaneous
%imbibition give global roughness
%exponent $\chi \approx 1.25$, indicating anomalous scaling.

One of the great advantages of the phase-field approach is that it's
possible to analytically derive equations of motion for the phase
boundaries in the so-called sharp interface limit \cite{Elder01}.
Most recently, we have developed a systematic formalism to derive
such equations for the 2D meniscus and 1D contact line dynamics of
fluids in capillary rise~\cite{majaniemi04}. The equations are
derived from the 3D bulk phase-field formulation,
%which can be deduced from a more microscopic density functional
%approach.
using variational approach as applied to relevant Rayleigh
dissipation and free energy functionals. Through successive
projections, equations of motion for the 2D meniscus and 1D contact
line can be derived. The leading terms of such equations (for small
amplitude, long wavelength fluctuations) can be shown to
agree with results obtained from the sharp interface equations in
the appropriate limits \cite{SamiPRL}.

In addition to the need for non-local models to account for mass
conservation, in Hele-Shaw and imbibition type of problems the
inherent quenched disorder should be properly taken care of. Unlike
thermal disorder, which is relatively easy to handle,
%usually renders good agreement between theories and simulations,
quenched disorder depends on the height of the 1D interface $h(x,t)$
as $\eta (x,h(x,t))$. This makes its influence on interface
roughening highly nontrivial, often leading to anomalous scaling
\cite{asikainen02,ramasco00}. Currently, for such cases good
agreement between theory, simulations and experiments has not been
achieved. Even on the experimental side some results such as the
quantitative values of the scaling exponents, are not consistent and
difficult to interpret. Very recently, Soriano {\it et
al.}~\cite{soriano02_03} conducted an experimental study of forced
fluid invasion in a specially designed Hele-Shaw cell. The quenched
disorder pattern in Hele-Shaw cell is realized by creating large
number of copper islands that randomly occupy the sites of a square
grid on a fiberglass substrate fixed on the bottom Hele-Shaw cell.
Three different disorder patterns were used. Two of them are
obtained by random selection of the sites of a square lattice. The
third kind of disorder is formed by parallel tracks, continuous in
the interface growth direction and randomly distributed along the
lateral direction. It was found that for forced flow the temporal
growth exponent $\beta \approx 0.5$ which is nearly independent of
experimental parameters and disorder patterns. However, the spatial
roughness exponent $\chi$ was found to be sensitive to experimental
parameters and disorder patterns. Anomalous scaling with $\chi
\approx 1.0$, and a local roughness exponent $\chi_{\rm local}
\approx 0.5$ was found in disorder pattern with parallel tracks
along the growth direction. It was also demonstrated that such
anomalous scaling is a consequence of different local velocities on
the tracks and the coupling in the motion between neighboring
tracks.

On the theoretical side, for non-local Hele-Shaw and paper
imbibition problems there are two different approaches within the
phase-field models to include additive quenched disorder. Dub{\'e} {\it
et al.}~\cite{Dube99_01,laurila05} put the quenched disorder inside
the chemical potential, the gradient of which is the driving force
for interface motion. On the other hand, Hernandez-Machado {\it et
al.}~\cite{hernandez-machado01} accounted for the effect of
fluctuation of Hele-Shaw gap thickness as a mobility with quenched
disorder in the phase-field model, while keeping the chemical
potential free of noise. These two approaches lead to quantitatively
different roughening properties.

When considering the problem of capillary rise in a typical
Hele-Shaw cell set-up between two rough walls more microscopically,
the location of the surface of such corrugated walls in the
Cartesian coordinate system is a spatially fluctuating quantity,
which indicates the presence of quenched disorder. An experimental
realization is given in \cite{geromichalos02}. To treat this problem
faithfully, in solving the phase-field equation such a fluctuating
wall surface should be treated as a physical boundary without
phenomenologically adding quenched noise to the equation of motion
as done previously. Consequently it is evident that a rigorous analytic
treatment of such a problem is overwhelmingly difficult. However, in
this paper we demonstrate that with proper mathematical formulation
of the problem, it is possible - albeit with some approximations -
to analytically derive equations of motion for the meniscus and
contact line dynamics. Most importantly, these equations incorporate
the wall disorder in a natural way. To achieve this, we utilize an
explicit curvilinear coordinate transformation of the phase-field
equation in order to apply projection methods to unravel the
relevant physics in the limit of small disorder. To some extent,
this kind of curvilinear coordinate transformation is similar to the
boundary-fitted coordinate system frequently used in Computational
Fluid Dynamics (CFD)~\cite{anderson95}.

The outline of this paper is as follows: In Chapter \ref{2D} we will
consider the phenomenological 2D phase field model of capillary rise
with quenched disorder in the mobility, similar to that in Refs.
\cite{hernandez-machado01,hernandez-machado03}. We will adapt the
systematic projection method of Kawasaki and Ohta \cite{kawasaki82}
to obtain a linearized interface equation (LIE) that describes small
fluctuations of an interface, whose deterministic part reduces to
the previous result \cite{hernandez-machado01,hernandez-machado03}
in a special limit. To treat the problem rigorously, In Chapter
\ref{3D} we will consider the full 3D phase field model with
corrugated walls as the source of quenched disorder. The
transformation to curvilinear coordinates, as discussed above, is
introduced to obtain linearized, effective bulk disorder from the
original curvilinear boundary condition. Following this we develop
and apply a general projection scheme \cite{majaniemi04} to obtain
the effective equations of motion for small fluctuations of the 2D
meniscus, and ultimately for the 1D contact line between the
meniscus and the wall. Again, the deterministic parts of these
equations reduce to previously known limits in special cases.
However, we demonstrate that the forms of the quenched noise terms
derived here are different from the previous works.

%Correspondence between the 1D contact line equation from the 3D model, and
%the LIE from the 2D model is found as expected. However, we find a qualitative
%difference in the effective noise terms, which disappears only in the case
%of columnar disorder (parallel tracks). Whether this difference is due
%to the different projection methods, or difference physics described by
%the 2D and 3D models, is unfortunately unknown at present. Actually, even
%the physical relevance of the difference, in terms of scaling exponents,
%is as of yet unknown.

%\begin{comment}
%In this paper, the above outlined scheme of dealing with the quenched
%disorder will be analyzed in the framework of phase-field models.
%Modification to the original Green's function is made as the direct
%consequence of curvilinear coordinate transformation. Successive
%projections to meniscus and contact line will be performed, which
%afford equations of motion for meniscus and contact line. It will be
%explicitly shown that the original quenched disorder on the wall
%surface will end up to a non-trivial quenched noise in the contact
%line equation. We study the annealed properties of this noise,
%which means that we will not take into account the coupling of
%interface fluctuations to the noise.
%\end{comment}

\section{2D Phase Field Model with Stochastic Mobility}
\label{2D}

A 2D phase-field model explicitly including the local conservation
of bulk mass was introduced to study capillary rise by Dub{\'e} \emph{et
al.} \cite{Dube99_01}. The bulk disorder in their model was included
in the effective chemical potential. Recently a similar model, where
the disorder was considered through a stochastic mobility
coefficient, was studied by Hernandez-Machado \emph{et al.}
\cite{hernandez-machado01,hernandez-machado03}. In particular, they
assumed a one-sided mobility coefficient, which vanishes on one side
of the interface. From this model they derived an equation of motion for small
interface fluctuations. In this section, we will use the systematic
projection method introduced by Kawasaki and Ohta \cite{kawasaki82},
to derive the corresponding linearized interface equation (LIE)
describing small fluctuations in a sharp interface in a similar
model. 
In our model we assume the mobility to be independent of the phase,
as in the previous works \cite{Dube99_01}, but spatially stochastic,
as in \cite{hernandez-machado01,hernandez-machado03}.
This corresponds to considering the invading fluid and the
porous medium, but not the receding fluid. This picture is valid
when the receding fluid has low density and viscosity. In practice
this would mean a gas, such as air, being displaced by a liquid, such
as water or oil. The model allows a systematic projection of the effective
noise term at the interface.

The phase field model describes capillary rise at a coarse-grained
level with a phase field $\phi(\mathbf{x,t})$ that obtains the value
$\phi=-1$ in the phase of the displaced fluid, and $\phi=+1$ in the
phase of the displacing fluid. The phase field thus describes the
effective component densities, and thus must be locally conserved.
An energy cost for an interface is included to obtain the free
energy as
\begin{equation}
\label{free energy}
\mathcal{F}[\phi(\mathbf{x},t)]=\frac{1}{2}[\nabla\phi(\mathbf{x,t})]^2
+ V(\phi(\mathbf{x,t})),
\end{equation}
where $V$ has two minima at $\phi=+1$ and $\phi=-1$. The details of
$V$ are not relevant in the sharp interface limit, except to define
the surface tension, so we can choose the standard Ginzburg-Landau
form $V(\phi)=-\phi^2/2+\phi^4/4-\alpha\phi$, where one of the
phases can be set metastable by nonzero coefficient $\alpha$. The
equation of motion for the conserved phase field is given by the
continuity equation $\partial_t\phi=-\nabla\cdot \mathbf{j}$ and
Fick's law $\mathbf{j}=-\tilde{M}\nabla\mu$, where
$\mu=\delta\cal{F}/\delta\phi$, and $\tilde{M}=M(1+\xi(x))$ is the
mobility that we choose to be a position dependent stochastic
variable here. The resulting equation of motion for the phase
field is then given by
\begin{equation}
\label{2Dpf}
\partial_t\phi(x,t)=\nabla\cdot \tilde{M}(x)\nabla \mu[\phi]
=M\nabla\cdot (1+\xi(x))\nabla\left[V'(\phi)-\nabla^2\phi\right],
\end{equation}
where the variable $\xi$ is now the dimensionless, quenched noise.
The sharp interface limit of this model without the noise is well
known, and discussed e.g. in ref. \cite{Elder01}.
The geometry of the problem is that of a half-plane, where a
reservoir of the displacing fluid is located at the $x$-axis. The
boundary condition of the chemical potential at the half-plane
boundary can be connected to the physical effect that is driving the
capillary rise. In this paper we will consider spontaneous
imbibition, where the rise is driven by a chemical potential
difference in the medium, which favors the displacing fluid
\cite{Dube99_01}. This means that the two minima of $V$ are at
different heights. In our notation the chemical potential difference
is $2\alpha$, and we consider chemically homogenous medium, where
$\alpha={const.}$. Spontaneous imbibition corresponds to a Dirichlet
boundary condition \hbox{($\mu=const.=0$)} at the reservoir
\cite{Dube99_01}. Forced flow, where flow is caused by an imposed
mass flux into the system from the reservoir, can be modeled with
the Neumann boundary condition \hbox{($\nabla\mu=F\hat{y}$)},
where $F$ is the flux \cite{laurila05}. An analysis along the lines
presented in this paper can also be conducted for the case of forced
imbibition. A recent review of phase field modeling of imbibition is
given in Refs. \cite{AlaNissila04,Alava04}.

Using the Green's function $G(r;r')$ for the 2D Laplacian,
equation \eqref{2Dpf} can be inverted using Gauss's theorem. This
leads to the integro-differential form
\begin{equation}
\begin{split}
\label{inverted 2d}
\frac{1}{M}\int_V dr'\sqrt{\det(g')} G\partial_t\phi'=(1+\xi)\mu
-\int_V dr'\sqrt{\det(g')}G \nabla'\xi'\cdot\nabla'\mu'\\
-\int_V dr'\sqrt{\det(g')}G\mu'\nabla'^2\xi'+\Lambda.
\end{split}
\end{equation}
%\int_S \vec{dS}'\cdot\left[G\nabla'\mu'-\nabla'G\mu'+G\xi'\nabla'\mu'-\nabla'G\xi'\mu'+G\nabla'\xi'\mu'\right].
%
Notation here has been shortened by omitting the function arguments,
and using unprimed and primed functions for functions of unprimed
and primed coordinates, respectively. The Green's functions always take
both primed and unprimed coordinates as argument.
Also the coordinate invariant
form is used, with integration measure given by $\sqrt{\det(g)}$. The boundary
term $\Lambda$ vanishes in the case of spontaneous imbibition, or
Dirichlet boundary condition in half-plane geometry.

Using the standard 1D kink solution method for projection to sharp
interface \cite{kawasaki82,bray94} in normal coordinates gives
Eq.~\eqref{inverted 2d} as
\begin{equation}
\label{yksvitunvalivaihe}
\begin{split}
\frac{1}{M}\int du\partial_u\phi_0\int ds'du'\sqrt{\det(g')} G\partial_t\phi'=
-(1+\xi|_{u=0})(\sigma\kappa+\int du\partial_u\phi_0\alpha)\\
+\int du\partial_u\phi_0\int ds'du'\sqrt{\det(g')}G\left[\partial_{u'}\xi'\partial_{u'}\mu'
+(1-2u'\kappa')\partial_{s'}\xi'\partial_{s'}\kappa'\partial_{u'}\phi_0'\right]\\
+\int du\partial_u\phi_0\int ds'du'\sqrt{\det(g')}G\nabla'^2\xi'(\kappa'\partial_{u'}\phi_0'+\alpha),
\end{split}
\end{equation}
where the normal coordinates $(s,u)$ are distances along and
perpendicular to the interface, respectively, $\kappa$ is the local
curvature of the interface, $\sigma=\frac{1}{2}\int du
\left(\partial_u\phi_0(u)\right)^2$ is the surface tension of the
phase field model, and finally $\phi_0$ is the 1D kink solution
$\partial_u^2\phi_0(u)=V'(\phi_0)$. In the Ginzburg-Landau form of
$V$ this would be given by $\phi_0(u)=\tanh(u/\sqrt{2})$, with the
appropriate choice of dimensionless units. We have assumed a
disorder correlation length that is larger than the interface width,
which leads to the constant surface tension obtained.
%This intermediate step is shown in detail, because
%a crucial approximation is made here.

With two further approximations \footnote{The term $\partial_u\mu$
is approximated with its form for the 1D disorder free system:
$\partial_u\mu\approx \frac{v_n}{M}\Theta(u)$, where $v_n$ is the
normal velocity. This approximation is similar to the 1D kink
solution in the standard procedure, but its range of extent is not
only near the interface, but between the reservoir and interface.
Another assumption we made in the derivation is neglecting boundary
terms of form $\int dx' G(x,H(x,t);x',0)\xi(x')$. This is justified
when the width of the Kernel $G$ along $x$-direction, which is of
order $H$, is much larger than the correlation length of the
disorder $\xi$.} the standard procedure \cite{bray94} can be followed.
Transforming the equation to Cartesian coordinates is made somewhat more 
tedious by the necessity to transform derivatives w.r.t. $s$ and $u$, but 
standard differential geometry methods can be applied.
After the sharp interface limit, i.e. $\phi_0 \rightarrow -1+2\Theta(u)$,
the transformation to Cartesian coordinates, and linearization in
small fluctuations of the interface $h$ and the noise $\xi$, which
also eliminates cross terms proportional to $h\xi$, we get the LIE as
\begin{equation}
\label{ImplicitLIE}
\begin{split}
\frac{1}{M}\int dx'\Big[G(x,H_0;x',H_0)+\partial_yG(x,y;x',H_0)\vert_{y=H_0}h(x,t)+\\
\partial_{y'}G(x,H_0;x',y')\vert_{y'=H_0}h(x',t)\Big]\partial_t\Big[H_0(t)+h(x',t)\Big]=\\
-\sigma\partial_x^2h(x,t)-\alpha+\frac{\partial_tH_0(t)}{M}\Xi(x,H_0(t)),
\end{split}
\end{equation}
where the disorder term $\Xi$ is given by
\begin{equation}
\label{hardnoise}
\Xi(x,y)=\int dx'\int_0^ydy'\xi(x',y')\partial_{y'}G(x,y;x',y').
\end{equation}
Note that the linearization has been carried out in full here. This
means that the disorder term does not include any dependence on the
interface fluctuations. This eliminates the non-linearity of the
quenched noise, which is one of its characteristic properties, but
we believe it is not crucial in the regime where the linearization
is appropriate. In other words, our results show non-trivial features
that arise in the effective noise at the interface level with this
type of multiplicative
bulk disorder, even in the linear regime of weak disorder.

The Green's function for the Dirichlet boundary condition in
half-plane geometry is explicitly given by
\begin{equation}
\label{2DGreen}
G(x,y;x',y')=
\frac{1}{4\pi}\ln{\frac{(x-x')^2+(y-y')^2}{(x-x')^2+(y+y')^2}}.
\end{equation}
Using this, the Fourier space representation of the interface
equation~\eqref{ImplicitLIE} becomes
\begin{equation}
\label{sm_nondriven_interface}
\begin{split}
\left(1-e^{-2\vert k\vert H_0(t)}\right)\partial_th(k,t)+
\vert k\vert\partial_tH_0(t)\left(1+e^{-2\vert k\vert H_0(t)}\right)h(k,t)=\\
-\sigma_B \vert k\vert^3 h(k,t)+\vert k\vert\partial_tH_0(t) \Xi(k,H_0(t))+|k|M\alpha_k,
\end{split}
\end{equation}
where $\sigma_B=M\sigma$, $\alpha_k$ is the Fourier transform of the
chemical potential difference ($\alpha_k=0$, if $k\neq 0$, when
$\alpha=const.$), and the disorder in Fourier space is given by
\begin{equation}
\label{nondriven_hardnoise}
\Xi(k,y)=-\frac{1}{2}\int_0^y dy' \xi(k,y')\left(e^{-\vert k\vert(y+y')}
+e^{-\vert k\vert(y-y')}\right).
\end{equation}
In the case of columnar disorder, which doesn't depend on $y$, the
interface equation simplifies to
\begin{equation}
\label{col_int_eom}
\partial_th(k,t)=-\frac{\vert k\vert\partial_tH_0(t)\left(1+e^{-2\vert k\vert H_0(t)}\right)
+\sigma_B \vert k\vert^3}{1-e^{-2\vert k\vert H_0(t)}} h(k,t)
+\partial_tH_0(t) \xi(k)+\frac{ \vert k \vert
M\alpha_k}{1-e^{-2\vert k\vert H_0(t)}}.
\end{equation}
It is noteworthy that the limit $k\rightarrow 0$ the interface
equation is $2H_0(t)\dot{H}_0(t)=\alpha_0$, readily giving the
correct Washburn law \cite{Dube99_01}, if we associate
$\lim_{k\rightarrow 0} h_k(t)=H_0(t)$, and $\lim_{k\rightarrow 0}
\alpha_k=\alpha_0$. \footnote{We note that these associations are
not consistent with the linearization in small disorder, since
different orders in an expansion should be kept separate. However,
the agreement between the long wavelength in linear order and the
zeroth order means that the same physical effect, namely mass
transport, both drives the propagation and dissipates long
wavelength fluctuations. This will not be the case when the flow is
forced, however.} Our method of analysis can be applied to the case
of forced flow by simply changing the boundary condition of the
phase field model at the reservoir, and applying the corresponding
Green's function \cite{laurila05}.

\begin{figure}
\includegraphics[width=8cm,height=10cm]{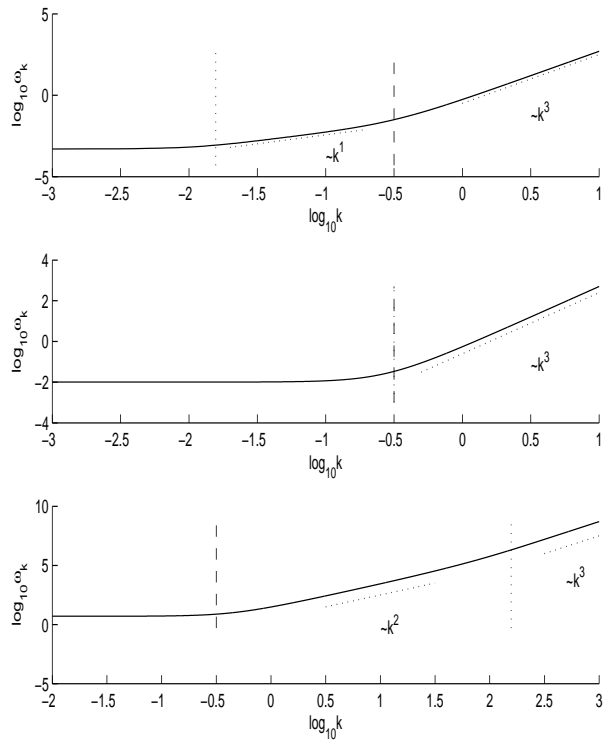}
\caption{The dispersion relation in Eqs.
\eqref{sm_nondriven_interface}, \eqref{corrug_menicus} and
\eqref{CL_result} in arbitrary units. The dispersion is determined
by the two length scales $\xi_x$ (vertical dashed line) and $H_0$
(vertical dotted line). The upper figure focuses on the ``deep'' regime,
with $H_0 \gg \xi_x$, in the middle figure these length scales are
the same, and the lowest figure focuses on the ``shallow'' regime, with
$\xi_x \gg H_0$.} \label{fig:dispersion}
\end{figure}

The dispersion relation \eqref{sm_nondriven_interface} above is the
main result in this section. It
involves two length scales: a crossover length scale
$\xi_x=2\pi\left(\frac{\sigma}{v}\right)^{\frac{1}{2}}$
\cite{Dube99_01}, and the distance from the reservoir $H_0$. The
deterministic part of the dispersion relation is plotted in Fig.
\ref{fig:dispersion}, at the two limits of these length scales: The
limit $H_0 \gg \xi_x$ brings out the ``deep'' limit, $kH_0 \gg 1$,
behavior. The limit $\xi_x \gg H_0$ shows the ``shallow'' limit,
$kH_0 \ll 1$, behavior.
A plot from the intermediate regime with $H_0 =
\xi_x$ is also shown.

The deterministic part of the dispersion relation here is identical
to that previously obtained by Dub{\'e} \emph{et al.} \cite{Dube99_01}
for the case of disorder in the chemical potential. In the ``deep''
limit where $\partial_th = -\left(\sigma_B \vert k \vert^3+\dot{H}_0
\vert k \vert \right)h$, our result also reproduces that of
Hernandez-Machado \emph{et al.} \cite{hernandez-machado01} for the
one-sided mobility case. Using different methods the same result has
also been obtained for the Hele-Shaw setup by Paune and Casademunt
\cite{paune03}, and Ganesan and Brenner \cite{ganesan98}.

Our disorder term in the LIE is similar to those obtained in Refs.
\cite{hernandez-machado01,paune03,ganesan98} in the sense that in
all cases the effective noise is linearly proportional to the
velocity of the interface propagation.  However, the quantitative
forms of the noise terms are different when using different methods.
How these differences influence the kinetic roughening of interfaces
would need to be determined by extensive numerical comparison
between the different results, which at this point has not been
conducted. As a linear $|k|$ proportionality in the Fourier space
representation of the effective noise term is linked to the $y$-derivative of
the Green's function of the Laplacian in real space representation, it
appears to us that the linear $|k|$ is present in the noise terms of
Refs. \cite{ganesan98} and \cite{paune03}, but not in Ref. \cite{hernandez-machado01}.
The linear $|k|$ dependence is in general characteristic of effective interface
noise caused by conserved bulk disorder \cite{Dube99_01}.
However, the $\vert k \vert $ dependence ($|k| \partial_t{H_0} \Xi(k, H_0(t))$)
dimensionally cancels the integral over the kernel in the
effective noise, Eq.\eqref{nondriven_hardnoise}. This is explicit in the
case of columnar disorder in Eq.\eqref{col_int_eom}, but is equally
valid with the noise in the non-columnar case.
Thus the multiplicative bulk disorder in the mobility
leads to a different type
of effective noise than the chemical potential disorder, which is
considered in \cite{Dube99_01}. Dimensionally this can be seen
from the definition of the model, Eq. \eqref{2Dpf}, where the noise
term is in front of the gradient of the chemical potential.

The fact that the columnar disorder leads to effective noise, which
is local in Fourier space, is in accordance with the conclusions
of experiments of Soriano \emph{et al.}
\cite{soriano02_03}, and with the numerical results from the
one-sided model \cite{hernandez-machado01}. This would indicate that
the phase dependence of the mobility is not crucially important when considering
the invasion of a viscous fluid into a fluid with negligible viscosity,
and when the interface is consequently always stable. When the direction of
the invasion is reversed, as studied recently with the one-sided model
in \cite{hernandez-machado03}, the situation is naturally quite different.

\section{3D Phase Field Model with Fluctuating Walls}
\label{3D}

While the stochastic mobility case of the previous section is
heuristically appealing, a more faithful treatment of the wall
disorder should start from the microscopic roughness of the walls.
To this end, in this chapter we will study a 3D version of the same
phase-field model as the 2D model, but where the mobility is constant
and the disorder is explicitly included as fluctuations of the wall
position. Thus the geometry of the model is that of a Hele-Shaw cell:
the 3D volume between two walls that are planar on average, but fluctuate.
We will show here that by proper mathematical formulation this
model can indeed be analyzed by a generalized projection formalism
\cite{majaniemi04}. The basic idea is to perform a mathematical
transformation from the basic Cartesian to a local curvilinear
coordinate system as defined by the wall itself. To this end, we
consider the one-wall setup as shown in Fig. 2. The one-wall setup
neglects the meniscus-mediated interaction between the two contact lines
at the two walls. The one-wall approximation also
neglects finite gap spacing, i.e. the distance between the two
walls in the Hele-Shaw cell, which fluctuates as result of the wall
fluctuations. This induces additional disorder effects when the wall
fluctuations are comparabale to the gap spacing, but it remains to be studied
if the gap effect can be separated from the contact line interaction,
which would be represented by two coupled equations of motion for the two
contact lines. In the present work, we only consider to the one-wall 
approximation, or the limit of large gap spacing.
%
%The meniscus between the walls is then considered as
%the superposition of the two independent stochastic processes at
%each wall.
%
%The gap spacing is a physical length scale that doesn't explicitly
%show up in the one-wall approach, but this obviously so in the 2D models
%as well. An analysis along the lines sketched here could be performed
%in the two-wall case, possibly yielding disorder terms due to fluctuations
%in the gap size, but this would be tedious to the extreme.
%
Disorder at the wall surface is taken into account by describing
local corrugations in the wall position around $y=0$ by a (small)
function $y=\delta H(x,z)$. The explicit coordinate transformation
to the local, curvilinear wall coordinate system is defined by
\begin{equation}
x' = x \,, \qquad y'=y - \delta H(x,z)\,,
\qquad z' = z\,,
\end{equation}
which corresponds to $y'=0$ when $y=\delta H(x,z)$. This means that
in the new coordinate system the wall is located back at $y'=0$.
Given the proper Green's function, ${\cal G}$ the phase field equation can be
inverted in any geometry and coordinate system as
\begin{equation}
\label{general_inverse}
\frac {\partial \phi} {\partial t} = M \nabla^2 \mu \ \ \Longrightarrow
\frac {1}{M} \int \upd V_1 {\cal G}(r,r_1) \frac {\partial \phi(r_1)}
{\partial t} =  \mu(r) + \Lambda_S\,,
\end{equation}
where $\Lambda_S$ is the corresponding surface term, and $dV_1$ is the
volume element for the coordinate system. The Green's
function appropriate for the above mentioned coordinate system is
considered in some detail in Appendix \ref{curvy}. Here we compute
the correction to the original Cartesian Green's function to linear
order in $\delta H(x,z)$. The final result we obtain, after
neglecting some surface contributions that are discussed in more
detail in Appendix \ref{curvy}, is what one would expect by simply
plugging the above definitions into the Cartesian Green's function
and linearizing in $\delta H$:
\begin{equation}
\label{fluctGreenMain}
\tilde{G}_{3D}(\boldsymbol{r}_1;\boldsymbol{r}_2)=G_{3D}(\boldsymbol{r}_1;\boldsymbol{r}_2)
-\delta H(\boldsymbol{r}_1)\partial_{y_1}G_{3D}(\boldsymbol{r}_1;\boldsymbol{r}_2)
-\delta H(\boldsymbol{r}_2)\partial_{y_2}G_{3D}(\boldsymbol{r}_1;\boldsymbol{r}_2).
\end{equation}
Here $G_{3D}$ is the Green's function for the Laplacian in 3D
Cartesian coordinates as given in Eq. \eqref{Lcart}. Here we again
only consider spontaneous capillary rise, where the boundary
conditions for the phase field model are zero chemical potential at
the reservoir and zero flux at the walls. Thus the surface integral
term in Eq.\eqref{general_inverse} is identically zero.

\subsection{Meniscus Dynamics}
\label{meniscus}

The projection and linearization of the integral equation follows
the standard projection operation theory
\cite{kawasaki82,zia85,bray94}, which we already used in the
previous chapter for the 2D model. The generalization for the
present case is straightforward. After projection the integral
equation is expressed in terms of the 2D meniscus variable $H(x,y)$
and has the following form:
\begin{equation}
\label{nonlin_meniscus}
\int \upd x' \upd y' \tilde {G}_{\textrm {3D}}
(x,y,H(x,y;t);x',y',H(x',y';t))
\frac {\partial H(x',y')} {\partial t} = \sigma_B \kappa.
\end{equation}
When linearizing the above equation, it must be done simultaneously
in the meniscus fluctuations, \emph{i.e.} $H(x,y;t)\simeq H_0(t) +
h(x,y;t)$, and in the wall fluctuations using the linearized Green's
function of Eq.\eqref{fluctGreenMain}. This results in the
linearized Green's function evaluated at the meniscus:
\begin{equation}
\begin{split}
\tilde{G}_{3D}(x,y,H(x,y);x',y',H(x',y'))\simeq G_{3D}(x,y,H_0;x',y',H_0)
\\+\delta H(x,H_0)\partial_yG_{3D}(x,y,H_0;x',y',H_0)
+\delta H(x',H_0)\partial_{y'}G_{3D}(x,y,H_0;x',y',H_0)\\
+h(x,y;t)\partial_zG_{3D}(x,y,z;x',y',H_0)|_{z=H_0}
+h(x',y';t)\partial_{z'}G_{3D}(x,y,H_0;x',y',z')|_{z'=H_0}.
\end{split}
\end{equation}
Substituting this into the meniscus equation \eqref{nonlin_meniscus} gives
\begin{eqnarray}
\label{meniscus raw eom}
I_{\textrm A } =  \sigma_B
\partial_y^2 H_0\,, \qquad
I_{\textrm B} + I_{\textrm C}
+ I_{\textrm D}+ I_{\textrm E}+ I_{\textrm F}  =
\sigma_B \nabla^2h(x,y;t)\,,
\end{eqnarray}
where the left hand side equation is to the zeroth order, and the
right hand side is to the first order in $h(x,y,t)$ or $\delta
H(x,H_0)$. These terms are defined in the same fashion as those in
Cartesian coordinate system~\cite{majaniemi04}. The terms
$I_{\textrm E}$ and $I_{\textrm F}$ arise from the fluctuating wall.
They are given by
\begin{eqnarray}
I_{\textrm A}(x,y) & \equiv &
\int \upd x_1 \int \upd y_1 G_{3D}
(x,y,H_0(t); x_1, y_1, H_0(t)) \partial_t H_0(t),
\\
I_{\textrm B}(x,y) & \equiv &
\int \upd x_1 \int \upd y_1  \partial_z {G}_{3D}
(x,y,z; x_1, y_1, H_0(t))|_{z=H_0}
h(x,y;t) \partial_t H_0(t) \,,
\\
I_{\textrm C}(x,y) & \equiv &
\int \upd x_1 \int \upd y_1  \partial_{z_1} {G}_{3D}
(x,y,H_0(t); x_1, y_1, z_1)|_{z_1=H_0}
h(x_1,y_1;t) \partial_t H_0(t) \,,
\\
I_{\textrm D}(x,y) & \equiv &
\int \upd x_1 \int \upd y_1 {G}_{3D}
(x,y,H_0(t); x_1, y_1, H_0(t)) \partial_t h(x_1,y_1;t) \,,
\\
I_{\textrm E}(x,y) & \equiv &
\int \upd x_1 \int \upd y_1  \partial_y {G}_{3D}
(x,y,H_0(t); x_1, y_1, H_0(t))
\delta H(x,H_0) \partial_t H_0(t) \,,
\\
I_{\textrm F}(x,y) & \equiv &
\int \upd x_1 \int \upd y_1  \partial_{y_1} {G}_{3D}
(x,y,H_0(t); x_1, y_1, H_0(t))
\delta H(x_1,H_0) \partial_t H_0(t) \,.
\end{eqnarray}
The zeroth order equation would give the Washburn law, if we used
the Green's function for the geometry between two walls and assumed
a constant curvature for the meniscus. 
We will assume an average profile $H_0(t)$, which can be considered to obey
Washburn's law even though we have only a single vertical
wall in the system. Since $H_0$ is not needed for determining
the form of the evolution equation for the fluctuating part
$h$ of Eq.\eqref{meniscus raw eom} at that single wall,
the precise time-dependence of $H_0$ is not
crucial for the analysis to be presented below. 

A local equation of motion for the meniscus fluctuations can be
obtained by Fourier-cosine transformation following
\cite{majaniemi04}. The above terms become
\begin{eqnarray}
{\mathcal{F}}_{x/k_x}{\mathcal{F}}^{cos}_{y/k_y}[I_{B}]&=&
\frac{1}{2}\dot{H}_0h(\vec{k},t);
\\
{\mathcal{F}}_{x/k_x}{\mathcal{F}}^{cos}_{y/k_y}[I_{C}]&=&
\frac{1}{2}e^{-2kH_0}\dot{H}_0h(\vec{k},t);
\\
{\mathcal{F}}_{x/k_x}{\mathcal{F}}^{cos}_{y/k_y}[I_{D}]&=&
\frac{1}{2k}\dot{h}(\vec{k},t)\left(1-e^{-2kH_0}\right);
\\
I_{\textrm E}&=&0;
\\
{\mathcal{F}}_{x/k_x}{\mathcal{F}}^{cos}_{y/k_y}[I_{\textrm F}]&=&\frac{\dot{H}_0(t)}{2k}
\left(1-e^{-2kH_0}\right)\delta H(k_x,H_0(t)),
\end{eqnarray}
where $k=\sqrt{k_x^2+k_y^2}$. We then have the meniscus equation of
motion using the above in the Fourier transform of Eq.
\eqref{meniscus raw eom}:
\begin{equation}
\label{corrug_menicus}
\partial_t h(\vec{k},t)=
-\frac{k\partial_t H_0(t)\left(1 + e^{-2H_0(t)k}\right)+
\sigma_Bk^3 } {\left(1 - e^{-2H_0(t)k}\right) } h(\vec{k},t)+
k\dot{H}_0\delta H(\vec{k},H_0).
\end{equation}
The deterministic part of the above meniscus equation is identical
to the deterministic part of the LIE derived from the 2D phase-field
model \eqref{sm_nondriven_interface}, apart from the dimensionality.
This is by construction, since the same method was used for the same
theory in different dimensions by applying the corresponding Green's
functions.

A similar analysis can also be performed for the case where the
disorder at the walls consists of chemical impurities (\emph{i.e.}
spatially fluctuating surface tension) instead of spatial roughness
\cite{majaniemi04}. In this case the deterministic part of the
meniscus equation is by construction identical to that of the above.
However, there is no effective noise at the meniscus level, since
the effect of the disorder comes in from the contact line that
serves as a boundary condition for the meniscus.

\subsection{Contact Line Dynamics}
\label{contact line}

To proceed to the level of the 1D contact line we
consider the generalized variational approach \cite{majaniemi04}.
Formally, one can write the 3D phase field model in
terms of variations of a Rayleigh dissipation functional, and a
free energy functional. Then, using approximations that express
higher dimensional entities in term of the relevant lower dimensional
ones, we obtain a chain of projection equations as
\begin{eqnarray}
\frac {\delta R_{\textrm {3D}}[ \dot {\phi}]}
{\delta \dot {\phi}(x,y,z;t)} & = & -
 \frac {\delta F_{\textrm {3D}}[\phi]}{ \delta \phi(x,y,z;t)}
\\
\label{2d_functional_eom}
\Rightarrow  \frac {\delta R_{\textrm {2D}}[ \dot H ] }
{\delta \dot H (x,y;t)} & = & -
 \frac {\delta F_{\textrm {2D}}[H]}{ \delta H(x,y;t)}
\\
\Rightarrow  \frac {\delta R_{\textrm {1D}}[ \dot C ] }
{\delta \dot C (x;t)} & = & -
 \frac {\delta F_{\textrm {1D}}[C]}{ \delta C(x;t)}\,,
\label{Ray}
\end{eqnarray}
where $R_{\textrm {dD}}$ refers to the Rayleigh dissipation
functional, and $F_{\textrm {dD}}$ refers to the free energy
functional in $d$ dimensional space. Here the relevant 3D, 2D and 1D
objects are the phase field, the meniscus profile and the contact
line profile, respectively. The variable $C(x;t)$ denotes the
fluctuating contact line profile, and $H(x,y;t) = H_0(t)+h(x,y;t)$
for the one-wall case. The corresponding expansion for the contact line
is $C(x,t)=C_0(t)+c(x,t)$. For small fluctuations $h$ and $c$, consistency
requires that $C_0(t)=H_0(t)$. The projection from 3D to 2D is made possible
by the 1D kink approximation in the direction normal to the
interface, as demonstrated in the preceding Section. The
corresponding approximation we have used to make the 2D to 1D
projection possible is the quasi-stationary (QS) approximation
$\nabla^2 h(x,y;t) = 0$ ; $h(x,0,t)=c(x,t)$, which corresponds to the minimum of energy
constrained by the contact line profile. The meniscus can then be
expressed in terms of the contact line as
\begin{eqnarray}
\label{quasistationary}
h_{\textrm {qs}}(x,y;t) &=& \int_{-\infty}^{\infty} \upd x_1\ \
g(x-x_1, y) c(x_1;t);\\
\label{qs_kernel}
g(k_{x},y) &=& e^{-|k_{x}|y} \qquad \Leftrightarrow \qquad g(x,y)=\frac{1}{\pi}\frac{y}{x^2-y^2}.
\end{eqnarray}
%
%We further discuss this approximation and possible extensions to it in the final chapter.

The explicit forms for the Rayleigh dissipation and free energy
functionals that reproduce the meniscus equation
\eqref{nonlin_meniscus} when plugged into
Eq.~\eqref{2d_functional_eom} are
\begin{eqnarray}
R_{2D}[\dot{H}]&=&\int_{-\infty}^{\infty}~dx_1dx_2\int_0^{\infty}~dy_1dy_2\int~dt_1
\dot{H}(x_1,y_1,t_1);\nonumber\\
&&\times \tilde{G}_{3D}(x_1,y_1,H(x_1,y_1,t_1);x_2,y_2,H(x_2,y_2,t))\dot{H}(x_2,y_2,t_1)
\label{2D_rayleigh}\\\hspace{2cm}
F_{2D}[H]&=&\sigma_B\int_{-\infty}^{\infty}~dx_1\int_0^{\infty}~dy_1\int~dt_1~\sqrt{1+|\nabla
H(x_1,y_1,t_1)|^2}.\label{2D_energy}
\end{eqnarray}
The effective 1D functionals can be obtained from the above by
inserting the quasi-stationary approximation $h_{\textrm {qs}}$ into
\eqref{2D_rayleigh} and \eqref{2D_energy}.
In order to obtain the
1D equation of motion to linear order in small fluctuations one needs to 
expand the functionals to second order in both $c(x,t)$ and $\delta H(x,y)$,
and then take the variation with respect to the contact line as shown in
Eq.\eqref{Ray}.

Neglecting the zeroth order equation for the reasons mentioned
earlier, the general Fourier space equation of motion we obtain for
the first order fluctuations is
\begin{equation}
{\mathcal{F}}_{x/k_x}\left[I_2+I_3+I_4+I_5+I_6\right]=-\sigma_B |k_x| c(k_x,t),
\end{equation}
where the LHS is the variation of the Rayleigh dissipation
functional, and the RHS is the variation of the free energy. The RHS
is recognized as the deterministic restoring force  acting on the
contact line \cite{joanny84}.
The shorthand notations stand for
\begin{eqnarray}
I_{2}(x)&=&2\dot{C}_0\int_{-\infty}^{\infty}dx_1dx_2dx_3\int_0^{\infty}dy_1dy_2~
g(x-x_1,y_1)\nonumber\\
&&\partial_{z_1}G_{3D}(x_1,y_1,z_1;x_2,y_2,C_0)|_{z_1=C_0}g(x_1-x_3,y_1)c(x_3,t)\\
I_{3}(x)&=&2\dot{C}_0\int_{-\infty}^{\infty}dx_1dx_2dx_3\int_0^{\infty}dy_1dy_2~
g(x-x_1,y_1)\nonumber\\
&&\partial_{z_2}G_{3D}(x_1,y_1,C_0;x_2,y_2,z_2)|_{z_2=C_0}g(x_2-x_3,y_2)c(x_3,t)\\
I_{4}(x)&=&2\int_{-\infty}^{\infty}dx_1dx_2dx_3\int_0^{\infty}dy_1dy_2~
g(x-x_1,y_1)\nonumber\\
&&G_{3D}(x_1,y_1,C_0;x_2,y_2,C_0)g(x_2-x_3,y_2)\dot{c}(x_3,t)\\
I_{5}(x)&=&2\dot{C}_0\int_{-\infty}^{\infty}dx_1dx_2\int_0^{\infty}dy_1dy_2~
g(x-x_1,y_1)\nonumber\\
&&\partial_{y_1}G_{3D}(x_1,y_1,C_0;x_2,y_2,C_0)\delta H(x_1,C_0)\\
I_{6}(x)&=&2\dot{C}_0\int_{-\infty}^{\infty}dx_1dx_2\int_0^{\infty}dy_1dy_2~
g(x-x_1,y_1)\nonumber\\
&&\partial_{y_2}G_{3D}(x_1,y_1,C_0;x_2,y_2,C_0)\delta H(x_2,C_0).\label{noise_realspace}
\end{eqnarray}
Not all of these integrals are solvable in closed form, but can be
approximated to a good degree of accuracy by the following
expressions:
\begin{eqnarray}
{\mathcal{F}}_{x/k_x}[I_{2}]&=&\frac{\dot{C}_0}{2|k_x|}c(k_x,t); 
\label{eq_3D_corrug_first_clEOM_term}\\
{\mathcal{F}}_{x/k_x}[I_{3}]&=&\frac{2\dot{C}_0}{\pi
|k_x|}c(k_x,t)\int_1^\infty ds
\frac{e^{-2|k_x|C_0s}}{s^3\sqrt{s^2-1}}\nonumber\\
&&\approx \frac{1.14\cdot 4}{3\pi
|k_x|}\dot{C}_0c(k_x,t)e^{-2.28|k_x|C_0}; \\
{\mathcal{F}}_{x/k_x}[I_{4}]&=&\frac{2\dot{c}(k_x,t)}{k_x^2\pi}\int_{1}^{\infty}ds
\frac{1- e^{-2|k_x|C_0s}}{s^4\sqrt{s^2-1}}\nonumber\\
&&\approx \frac{4\dot{c}(k_x,t)}{3\pi k_x^2}
\left(1-e^{-2.28|k_x|C_0}\right);\\
{\mathcal{F}}_{x/k_x}[I_{5}]&=&0;\\
{\mathcal{F}}_{x/k_x}[I_{6}]&=&\frac{2}{\pi |k_x|}\dot{C}_0\delta
H(k_x,C_0) \int_1^\infty ds \frac{1-
e^{-2|k_x|C_0s}}{s^2\sqrt{s^2-1}}\nonumber\\
&&\approx\frac{2}{\pi |k_x|}\dot{C}_0\delta H(k_x,C_0) \left(1-
e^{-2.56|k_x|C_0}\right).
\end{eqnarray}
We note that the corrections to the free energy functional in the
curvilinear coordinates are of third order in $\delta H$ and $h$.
This can be seen by coordinate transforming the area element, which
in Cartesian coordinates is $\sqrt{1+(\nabla h)^2}\approx
1+\frac{1}{2}(\nabla h)^2$.

Finally, after approximating $\frac{1.14\cdot4}{3\pi}\approx
\frac{1}{2}$ and
$\frac{\left(1-e^{-2.56|k_x|C_0}\right)}{\left(1-{e}^{-2.28|k_x|C_0}\right)}\approx{1}$,
the equation of motion for the contact line fluctuations becomes
\begin{equation}
\label{CL_result}
\dot{c}(k_x,t)=-\frac{\frac{3\pi |k_x|\dot{C}_0}{8}\left(1+
e^{-2.28|k_x|C_0}\right)+\sigma_B|k_x|^3}
{\left(1-{e}^{-2.28|k_x|C_0}\right)}c(k_x,t)+
\frac{3}{2}|k_x|\dot{C}_0\delta H(k_x,C_0).
\end{equation}
Note that all of the approximations above are for dimensionless
quantities, with errors depending on the physical parameter
$|k_x|C_0$. The relative errors in the approximated functional forms are under 3\%,
when compared against numerical integration of the respective integrals
for different values of $|k_xC_0|$. An exception to this are relative
errors of ${\mathcal{F}}_{x/k_x}[I_{4}] $ and ${\mathcal{F}}_{x/k_x}[I_{6}]$
when $|k_x|\rightarrow 0$, as both $I_{4}$ and $I_{6}$ vanish at the limit, 
causing the relative error behave badly. However, at machine precision away
from $|k_x|=0$ then these errors are no more than 15\%, and more importantly
the error of the complete dispersion relation stays within the 3\% error margin.
This is due to the fact that the dispersion remains finite, as one can see from
Figure \ref{fig:dispersion}.

Apart from simple numerical factors, the contact line equation above
has the same functional form as the results derived in the previous
sections. In particular, $\partial_th = -\left(\sigma \vert k_x
\vert^3+\dot{H}_0 \vert k_x \vert \right)h$ in the ``deep'' limit
$k_x^{-1} \ll H_0(t)$, which thus agrees with the previous works
discussed earlier \cite{hernandez-machado01,paune03,ganesan98}. This
form of dispersion relation is always obtained by our method for
interfaces in Model B. This has to do with the quasi-stationary
approximation, which essentially assumes that meniscus fluctuations
dampen quickly in the direction perpendicular to the contact line,
in order to obtain temporally local equations. How this leads to the
coupling of the meniscus and contact line dynamics is discussed in
more detail in another publication \cite{SamiPRL}.

The effective noise term we obtain from the 3D model shares the
property of linear dependence on the velocity of the propagation
with the 2D mobility noise, and with the previous analyses
\cite{hernandez-machado01,paune03,ganesan98}. In the case of surface
tension impurities at the wall \cite{majaniemi04} the effective contact line
noise is proportional to $k_x^2$, whereas in the
fluctuating wall case in Eq.\eqref{CL_result} the $|k_x|$
dependence is linear. This is analogous to the 2D mobility disorder
in the sense that the effective noise is different from that obtained
for conserved disorder. The
extra factor of $\vert k_x \vert$ as compared to the 2D model
(Eq.\eqref{sm_nondriven_interface}) comes
from the fact that the disorder in the 3D model comes from the
walls, whereas in the 2D model the disorder is in the bulk. We note
that the more complicated properties of the noise in the form of
non-locality in Fourier space are lost by our approximations. Note
that the noise is still non-local in real space, as is apparent from
its real space representation $I_6$ in Eq. \eqref{noise_realspace}.
In addition to Fourier space non-localities, our one-wall approach
doesn't explicitly include the gap spacing, which
provides an additional physical length scale
\footnote{By using the Green's function of the two-wall geometry it
would be a possible but an extremely lengthy calculation to take the
gap spacing explicitly into account in our model, but the result
would be two coupled equations of motion.}.
In spite of this, our results are
in accordance with those of Ref. \cite{paune03}, where the
gap fluctuations were considered as the only source of disorder
in context of Darcy's law.

\section{Discussion and Conclusions}
\label{discussion}

In this paper, we have studied the effective interface dynamics of
the three-phase contact line in a Hele-Shaw experiment by deriving
the meniscus and contact line equations of motion from higher
dimensional bulk phase-field theories by projection methods. The
projection methods take into account the non-local dynamics of the
system caused by local mass conservation, and can be systematically
applied starting from a full 3D description. We have here considered
two particular models, namely an \emph{ad hoc} model where the
disorder is in the effective mobility in 2D
\cite{hernandez-machado03}, and a more microscopic model where wall
corrugation in 3D is explicitly treated with a curvilinear
coordinate transformation. In both cases, we have focused on the
limit of small disorder by linearizing in disorder strength and in
the fluctuations caused by the disorder. By construction this
linearization, performed in real space, causes the Fourier space
representations of the equations of motion to be local. The upside
of this is that the effective dynamics are written in a concise
manner, and the physical predictions are easily interpreted and the
equations we obtain are readily affable to numerical analysis. The
obvious downside is that the procedure involves a number of
approximations, the validity of which is not certain \emph{a
priori}.

In particular, the quasi-stationary approximation of Eq.
\eqref{quasistationary}, which ultimately enables our contact line
analysis, requires a critical assessment. A more rigorous approach
would in fact consider the contact line as the boundary condition to
the meniscus equation of motion \eqref{corrug_menicus}. However,
explicitly solving the meniscus profile as a function of the contact
line leads to an equation that is, among other things, non-local in
time. Thus we are forced to simplify the model by using the QS
approximation, the validity of which we can consider both from a
physical perspective, or more rigorously by considering the limits
of the meniscus equation of motion. Physically, the QS approximation
comes from the minimum of meniscus energy constrained by the
boundary condition of the contact line. This is expected to define
the meniscus profile when the meniscus moves slowly, and thus it's
called the quasi stationary approximation. Mathematically the
meniscus equation \eqref{corrug_menicus} reduces to the diffusion equation when $C_0 \ll
k^{-1} \ll \sqrt{{\sigma}/{\dot{C}_0}}$, and
$\partial_th(x,y,t)\approx 0$. In this limit the meniscus
level disorder $I_F$ acts as a source term
\begin{equation}
\sigma_B\nabla^2h(x,y;t)=I_{\textrm F}.
\end{equation}
This leads to an additional disorder term in the QS approximation,
which then leads to a plethora of new first order disorder terms in
$R_{1D}$ and $F_{1D}$. However, all these new disorder terms arising
from $R_{1D}$ are proportional to $\dot{C}_0^2$, and thus not
relevant in the QS limit. Additionally, the two new disorder terms
created in $F_{1D}$ due to $I_F$ cancel each other out exactly. Thus, we expect
our results with the simplified version of the QS approximation to
hold in this particular limit.

In addition to the detailed derivations and new projection formalism
presented here, our purpose has been to quantitatively compare two
different approaches to modeling rough wall Hele-Shaw experiments,
namely that based on the 2D phase field model with a stochastic
mobility, Eq.\eqref{2Dpf}, and the 3D phase field model with a
fluctuating geometry. The projection method we use for both cases
produces the linear response of the meniscus and contact line to
small fluctuations. For both cases, the $k$ dependence of the
meniscus and contact line deterministic LIEs is the same.
%
%Comparing with other related works,
%instabilities have been shown to arise when the mobility
%in the phase field model is made to be phase-dependent
%in Ref. \cite{HLMP}. We do not find such behavior here as in our case the interface remains
%linearly stable.
%
In particular, in the special case of the ``deep'' limit where
$k^{-1} \ll H_0(t)$, the asymptotic forms of our general dispersion
relations are in agreement with previous works on the Hele-Shaw
problem by Paune and Casademunt \cite{paune03}, Ganesan and Brenner
\cite{ganesan98} and Hernandez-Machado \emph{et al.}
\cite{hernandez-machado01}. The main advantage of our method is the
way it incorporates the noise into the projection, and thus allows
us to study the effective noise caused to the contact line level by
bulk or wall disorder. The main result of this analysis is that in
both cases the effective noise is linearly proportional to the
velocity of the interface. While this result qualitatively agrees
with the other works cited above, quantitative differences remain in
the form of the noise terms. The relevance of these differences to
the actual kinetic roughening of the interfaces remains a
challenging numerical problem.

%A sharp interface calculation for the 3D model setup, where the reservoir
%has been taken into account, has also been done \cite{KenPRL}, and our
%results concerning the dissipation agree in the appropriate limit,
%as the sharp interface results are more general when dissipation is concerned.

\section{Acknowledgments:}

We would like to thank K. Elder for inspirational and pleasant
discussions, as well as for sharing his expertise and results on the
sharp interface analysis. This work has been supported in part by
the Academy of Finland through its Center of Excellence grant
(COMP). S.M. has been supported by a personal grant from the
Academy of Finland.

\section*{Appendix}
\appendix
\section{Curvilinear Coordinate System by Fluctuating Wall}

\label{curvy}
\begin{figure}
\includegraphics[width=12cm,height=8cm]{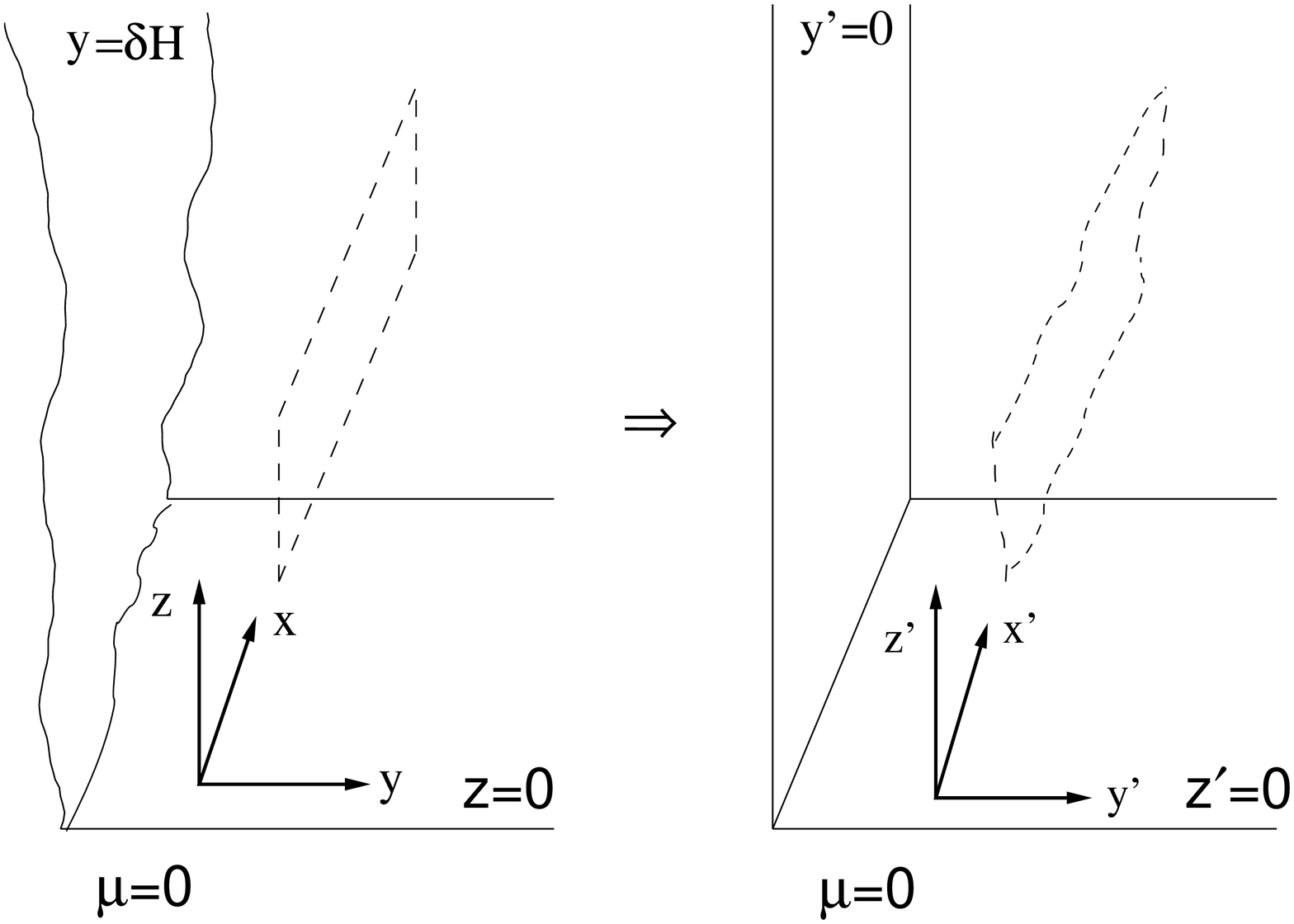}
\caption{A schematic presentation of the curvilinear coordinates
considered in the appendix. When presented in terms of
curvilinear coordinates, the rough wall by definition look straight.
However, the shift has introduced a coordinate fluctuation in space,
where a previously rectangular object looks curved when presented in terms
of the new coordinates. This gives rise to a bulk representation of
the fluctuating wall, which can be analyzed more easily than the
original boundary condition representation.} \label{schem}
\end{figure}

In this Appendix we will consider in more detail the Green's
function of the Laplacian in the fluctuating coordinate system
\begin{equation}
x' = x \,, \qquad y'=y - \delta H(x,z)\,,
\qquad z' = z\,.
\end{equation}
which is schematically depicted in Figure \ref{schem}. 
The generalization to the two-wall setup is straightforward, but
very tedious including two independent disorder functions.
In particular, we will consider the correction to the
Green's function to linear order in small wall fluctuations $\delta H(x,z)$.
First, the metric tensor of the above coordinate system can be obtained by
transformation of the Cartesian metric tensor as
\begin{equation}
\left[g_{i'j'}\right]=\sum_i \frac{\partial x^{i}}{\partial x^{i'}}
\frac{\partial x^{i}}{\partial x^{j'}}=
\begin{bmatrix}
1+(\partial_x\delta H)^2 & \partial_x\delta H &\partial_z\delta
H\partial_x\delta H\\
\partial_x\delta H & 1 & \partial_z\delta H\\
\partial_z\delta H\partial_x\delta H & \partial_z\delta H & 1+(\partial_z\delta
H)^2
\end{bmatrix}.
\end{equation}
The coordinate transformation from Cartesian coordinates is merely a
shift, and thus the integration measure should not change, meaning
that the Jacobian in the coordinate transformation of an integral
should be unity. This is indeed so, since
\begin{equation}
\det([g_{i',j'}])\equiv 1.
\end{equation}
In the case of Cartesian coordinates we can obtain the Green's
function, which we denote $G_{\textrm {3D}}$, using the image charge
method with the Dirichlet boundary condition at $z =0$ and the
Neumann boundary condition at $ y=0$:
\begin{equation}
\label{Lcart}
G_{\textrm {3D}}  =   G_{\textrm 3D}^+
+ G_{\textrm {3D}}^-\,,
\end{equation}
\begin{equation}
G_{\textrm {3D}}^{\pm}   =
  \frac {1}{4 \pi} \left[
\frac {1} {\sqrt {(x-x_1)^2 + (y \pm y_1)^2 + (z-z_1)^2}}
 - \frac {1} {\sqrt {(x-x_1)^2 + (y \pm y_1)^2 + (z+z_1)^2}} \right].
\end{equation}
We work in the limit of small fluctuations, so we write the
Laplacian in the curvilinear coordinates as the Cartesian Laplacian
plus a correction, $\tilde{\nabla}^2=\nabla^2 + L$. Note that we
unconventionally denote
$\nabla^2=\partial_{x_1}^2+\partial_{x_2}^2+\partial_{x_3}^2$ for
any coordinates $[x_1, x_2, x_3]$. The correction $L$ is explicitly
shown below:
\begin{eqnarray}
L= 2 \left[ \frac {\partial \delta H(x,z)}
{\partial x} \right]
\frac { \partial^2}{\partial x \partial y}
+2 \left[ \frac {\partial \delta H(x,z)}
{\partial z} \right]
\frac { \partial^2}{\partial z \partial y}
 + [ \frac {\partial^2 \delta H(x,z)}
{\partial x^2} +  \frac {\partial^2 \delta H(x,z)}
{\partial z^2} ] \frac {\partial} {\partial y}.
\label{CorrectionOperator}
\end{eqnarray}
In order to use Eq.\eqref{general_inverse} for the curvilinear
coordinates, we need the Green's function, which has the  property
of $\left[ \nabla'^2 + L' \right] \tilde {G}_{\textrm {3D}}
(\boldsymbol {r}',\boldsymbol {r_1}') =  - \delta (\boldsymbol {r}'
- \boldsymbol {r_1}')$. Since the Laplacian in the curvilinear
coordinates can be expressed as the Cartesian Laplacian plus a
correction, we can find the inverse of the curvilinear Laplacian, or
$\tilde{G}_{\textrm {3D}}$, to first order in $\delta H$ as
\begin{equation}
\tilde {G}_{\textrm {3D}}
(\boldsymbol {r}', \boldsymbol {r_1}')   \cong
\left( \nabla'^2 \right)^{-1} -  \left( \nabla'^2 \right)^{-1}
L'  \left( \nabla'^2 \right)^{-1}\,,
\end{equation}
The above operator notation can be written in full form as:
\begin{equation}
\tilde {G}_{\textrm {3D}}
(\boldsymbol{r}';\boldsymbol{r}'_1)
\approx G_{\textrm {3D}} (\boldsymbol{r}';\boldsymbol{r}'_1) -
 \int_{- \infty}^{\infty} \upd x_2  \int_{0}^{\infty} \upd y_2
 \int_{0}^{\infty} \upd z_2
G_{\textrm {3D}}
(\boldsymbol{r}';\boldsymbol{r}'_2) L(\boldsymbol{r}'_2)
 G_{\textrm {3D}}(\boldsymbol{r}'_2;\boldsymbol{r}'_1).
\label{Green1}
\end{equation}
Substituting $G_{\textrm {3D}}$ into the above, we can work out the
correction to the Green's function. After using a similar argument to
neglect surface integrals of type $\int dx' G(x,H(x,t);x',0)\xi(x')$
as we did with Eq.\eqref{yksvitunvalivaihe}, we find
\begin{equation}
\label{fluctGreen}
\tilde{G}_{3D}(\boldsymbol{r}'_1;\boldsymbol{r}'_2)=G_{3D}(\boldsymbol{r}'_1;\boldsymbol{r}'_2)
-\delta H(\boldsymbol{r}'_1)\partial_{y'_1}G_{3D}(\boldsymbol{r}'_1;\boldsymbol{r}'_2)
-\delta H(\boldsymbol{r}'_2)\partial_{y'_2}G_{3D}(\boldsymbol{r}'_1;\boldsymbol{r}'_2).
\end{equation}
At this point the primes can just be dropped, since $\partial_y=\partial_{y'}$.
This result is hardly surprising, since a simple substitution of $
y'=y - \delta H(x,z)$ to
$G_{3D}(\boldsymbol{r}'_1,\boldsymbol{r}'_2)$ yields identical
results to linear order.

The neglected surface integrals include a reservoir term and a wall
term. The reservoir term can be readily seen to be small when the
meniscus is further away from the reservoir than the disorder
correlation length. Additionally, the reservoir boundary correction
is zero when considering columnar \emph{i.e.} $z$-independent
disorder. The wall term is more problematic, since it involves the
boundary correction due to fluctuation in the direction of the wall
normal. We have to observe the meniscus further away from the wall
than the disorder correlation length in order to neglect this
boundary correction. The absence of boundary disorder corrections is
highly desirable if we are to keep our formalism tractable, so we
have neglected the boundary corrections to the Green's function.

\renewcommand\baselinestretch{1.08}\large

% -- CUT HERE -- -- CUT HERE -- -- CUT HERE -- -- CUT HERE -- %


\normalsize
\begin{thebibliography}{0}

\bibitem{moldovan00} D. Moldovan and L. Golubovic,
{\it {Phys. Rev. E}} {\bf 61} 6190 (2000).

\bibitem{bouchaud90} J. P. Bouchaud and A. Georges,
{\it {Phys. Rep.}} {\bf 195}, 127 (1990).

\bibitem{scheidegger57} A. E. Scheidegger, {\it The Physics of Flow through Porous
Media} (MacMillan Co., New York, 1957).

\bibitem{krug97} J. Krug, {\it Adv. Phys.} {\bf 46}, 139 (1997).

\bibitem{moulinet04} S. Moulinet, A. Rosso, W. Krauth, and E. Rolley,
{\it Phys. Rev. E} {\bf 69}, 035103(R) (2004).

\bibitem{joanny90} J. F. Joanny and M. O. Robbins,
{\it J. Chem. Phys.} {\bf 92}, 3206 (1990).

\bibitem{ertas94} D. Ertas and M. Kardar,
{\it Phys. Rev. E} {\bf 49}, R2532 (1994).

\bibitem{barabasi95} A.-L. Barab\'asi and H. E. Stanley, {\it Fractal Concepts in
Surface Growth} (Cambridge 1995).

\bibitem{paperburning}
J. Maunuksela et al. {\it Phys. Rev. Lett.} {\bf 79}, 1515 (1997);
M. Myllys et al. {\it Phys. Rev. Lett.} {\bf 84}, 1946 (2000);
M. Myllys et al. {\it Phys. Rev. E} {\bf 64}, 036101 (2001).

%\bibitem{family85} F. Family and T. Viscek, {\it J. Phys. A}; {\bf 18} (1985) L75.

%weak disorder
\bibitem{soriano02_03} J. Soriano, J. Ortin and A. Hernandez-Machado,
{\it Phys. Rev. E}, {\bf 66} 031603 (2002).
%strong disorder
J. Soriano, J. Ortin and A. Hernandez-Machado,
{\it Phys. Rev. E}, {\bf 67} 056308 (2003).

\bibitem{paune03} E. Paune and J. Casademunt, {\it Phys. Rev. Lett.},
{\bf 90} 144504 (2003).

\bibitem{geromichalos02} D. Geromichalos, F. Mugele and S. Herminghaus,
{\it Phys. Rev. Lett.}, {\bf 89} 104503 (2002).

\bibitem{buldryev92} S. V. Buldyrev {\it et al.}, {\it Phys. Rev. A}, {\bf
45} R8313 (1992).

\bibitem{horvath95} V.K. Horvath and H. E. Stanley,
{\it Phys. Rev. E}, {\bf 52} 5166 (1995).

\bibitem{amaral94} L. A. N. Amaral {\it et al.}, {\it Phys. Rev. Lett.},
{\bf 72} 641 (1994).

\bibitem{Dube99_01} M. Dub\'e {\it et al.},
{\it Phys. Rev. Lett.} {\bf 83}, 1628 (1999);
{\it Europ. Phys. Journal B} {\bf 15}, 701 (2000);
{\it Phys. Rev. E} {\bf 64}, 051605 (2001).

\bibitem{laurila05} T. Laurila, C. Tong, I. Huopaniemi, S. Majaniemi and T. Ala-Nissila,
{\it Europ. Phys. Journal B} {\bf 46} 553 (2005).

\bibitem{Elder01} K. R. Elder, M. Grant, N. Provatas and J. M. Kosterlitz,
{\it Phys. Rev. E} {\bf 64}, 021604 (2001).

\bibitem{majaniemi04} S. Majaniemi, Ph. D. Thesis, Helsinki University of Technology,
Finland (2004).

\bibitem{SamiPRL} S. Majaniemi, K. R. Elder, C. Tong, T. Laurila and T. Ala-Nissila,
submitted to Phys. Rev. Lett.

\bibitem{asikainen02} J. Asikainen, S. Majaniemi, M. Dub{\'e}, J. Heinonen and
T. Ala-Nissila, {\it Europ. Phys. Journal B} {\bf 30}, 253 (2002).

\bibitem{ramasco00} J. J. Ramasco, J. M. Lopez, M. A. Rodriguez,
{\it Phys. Rev. Lett.} {\bf 84}, 2199 (2000).

\bibitem{hernandez-machado01} A. Hernandez-Machado, J. Soriano, A. M. Lacasta,
M. A. Rodriguez, L. Ramirez-Piscina and J. Ortin,
{\it Europhys. Lett.}, {\bf 55} 194 (2001).

\bibitem{anderson95} J. D. Anderson Jr., {\it Computational Fluid
Dynamics: The Basics with Applications} (McGraw-Hill, New York, 1995)

%fingers with one-sided model
\bibitem{hernandez-machado03} A. Hernandez-Machado, A. M. Lacasta, E. Mayoral,
and E. C. Poire, {\it Phys. Rev. E}, {\bf 68} 046310 (2003).

\bibitem{kawasaki82} K.\ Kawasaki, T.\ Ohta, Prog. Theo. Phys. {\bf 68},
                      129 (1982).

\bibitem{AlaNissila04} T. Ala-Nissila, S. Majaniemi and K. Elder,
{\it Lect. Notes Phys.} {\bf 640}, 357 (2004).

\bibitem{Alava04} M. Alava, M. Dub{\'e} and M. Rost,
{\it Adv. Phys.} {\bf 113} 83 (2004).

\bibitem{bray94} A. J.\ Bray, Adv.\ in Phys.\ {\bf 43}, 357 (1994).

\bibitem{ganesan98} V. Ganesan and H. Brenner
{\it Phys. Rev. Lett.}, {\bf 81} 578 (1998).

\bibitem{zia85} R. K. P. Zia, {\it Nucl. Phys. B}, {\bf 251} 676 (1985).

%\bibitem{barton89} G. Barton, {\it Elements of Green's Functions and
%Propagation} (Clarendon Press, Oxford, 1989).

\bibitem{joanny84} J.F. Joanny and P.G. de Gennes,
{\it J. Chem. Phys.} {\bf 81} 552 (1984)

\end{thebibliography}
\end{document}